\documentclass[11pt]{article}
\usepackage{acl2015}
\usepackage{times}
\usepackage{url}
\usepackage{latexsym}
\usepackage{graphicx}
\usepackage{subcaption}
\usepackage{natbib}
\usepackage{array}

\title{Using LLM-Based Approaches to Enhance and Automate Topic Labeling}

\author{Trishia Khandelwal \\
  {\tt trishiakhandelwal007@gmail.com}}

\date{}
\begin{document}
\maketitle
\begin{abstract}
Topic modeling has become a crucial method for analyzing text data, particularly for extracting meaningful insights from large collections of documents. However, the output of these models typically consists of lists of keywords that require manual interpretation for precise labeling. This study explores the use of Large Language Models (LLMs) to automate and enhance topic labeling by generating more meaningful and contextually appropriate labels. After applying BERTopic for topic modeling, we explore different approaches to select keywords and document summaries within each topic, which are then fed into an LLM to generate labels. Each approach prioritizes different aspects, such as dominant themes or diversity, to assess their impact on label quality. Additionally, recognizing the lack of quantitative methods for evaluating topic labels, we propose a novel metric that measures how semantically representative a label is of all documents within a topic. 
\end{abstract}

\section{Introduction}

Topic modeling has emerged as a useful method for categorizing and making sense of the vast amounts of text data generated across various fields. Advances in topic modeling methods, such as BERTopic \citep{grootendorst2022}, have improved the coherence and interpretability of topics by leveraging embeddings that capture both context and semantics. However, the output of these models typically consists of keyword lists that reveal underlying themes, offering a preliminary understanding of the data, which alone are not very intuitive or sufficient for full interpretation. To derive actionable insights, these keyword-based topics often require manual insights by domain experts \citep{rijcken2023}, a process that is both time-intensive and inefficient for large-scale applications.

Automating the creation of topic labels is crucial, as these labels provide concise, meaningful summaries that enhance the depth of analysis and improve the interpretability of the topics. Traditionally, this labeling has been a manual process, posing a significant challenge when dealing with the complexity and volume of modern datasets. In this paper, we explore the potential of Large Language Models (LLMs) to automate topic labeling. We propose four approaches that utilize LLMs' ability to generate accurate, contextually rich labels. LLMs excel at interpreting and generating text while processing large amounts of data, making them well-suited to the task of automating topic labeling in a scalable and efficient manner.

Additionally, we introduce a novel metric for evaluating the representativeness of topic labels in relation to all documents within a topic. By leveraging embeddings and C\textsubscript{V} coherence, we calculate the pairwise coherence between topic labels and the corresponding documents. Typically, the effectiveness of topic labeling methods is evaluated manually, which can result in inconsistent assessments across different experiments. Our metric offers a more standardized approach for evaluating topic labeling performance.

The key contributions of this research are:

\begin{enumerate}
    \item Investigating and leveraging LLMs to generate more interpretable topic labels.
    \item Combining individual LLM-generated document summaries with top keywords to create richer topic labels.
    \item Developing a novel metric for quantitatively assessing the performance of topic labeling methods.
\end{enumerate}

\section{Related Work}

Research in the field of topic naming has evolved significantly over the years, starting from early probabilistic and bag-of-words approaches to more recent applications of deep learning and contextual embeddings.

Mei et al. \citep{mei2007} used probabilistic methods for automatic labeling, looking at the problem as an optimization task to minimize the Kullback-Leibler divergence between topic word distributions and label word distributions. Aletras and Stevenson \citep{aletras2014} proposed a graph-based approach, generating topic labels by querying topic keywords and constructing a graph from the resulting search words, which are then ranked. Lau et al. \citep{lau2011} used top-ranking topic keywords to generate a candidate label set from Wikipedia titles, which was then ranked to identify the best label based on association measures and lexical features.

Wan and Wang \citep{wan2016} utilized text summarization techniques to generate topic labels. Their method applied submodular optimization to extract sentences from the most relevant documents within a topic, summaries with high relevance and coverage. More recently, Alokaili et al. \citep{alokaili2020} trained a sequence-to-sequence neural model on a synthetic dataset derived from Wikipedia to automate topic labeling. The performance of their approach was evaluated using BERTScore, which measures the similarity between the generated labels and the gold standard labels \citep{zhang2020}.

The development of Large Language Models (LLMs) marked a major shift in labeling approaches. LLMs can process vast amounts of text data and are highly effective in capturing semantic meaning and identifying similarities in documents to generate meaningful topic labels. Kozlowski et al. applied BERTopic \citep{grootendorst2022} to extract the top 10 words for each topic, and then used the Flan, GPT-4o, and GPT-4 mini models to generate topic labels from these keywords \citep{kozlowski2024}. The accuracy of the labels in representing the topics was evaluated qualitatively.

\newcolumntype{C}[1]{>{\centering\let\newline\\\arraybackslash\hspace{0pt}}m{#1}}

\begin{table*}
    \centering
    \scalebox{1.35}{
        \begin{tabular}{|C{1in}|C{1in}|C{1in}|} \hline   
            \centering \textbf{Approach}& \centering \textbf{BBC}& \textbf{20 Newsgroup} \\ \hline
            \centering 1& \centering 0.0989 &  0.1183\\\hline
            \centering 2& \centering 0.0665 & \textbf{0.1324} \\ \hline  
            \centering 3& \centering \textbf{0.1200} & 0.1295 \\\hline
            \centering 4& \centering 0.0853& 0.1202 \\\hline 
        \end{tabular}
    }
    \caption{The values calculated by our metric for the various approaches described above on the BBC and 20 Newsgroups dataset.}
    \label{tab:my_label1}
\end{table*}

\section{Methodology}

\subsection{Topic Modeling}

We must perform topic modeling on data before we generate labels. We do so by incorporating the GPT-3.5-Turbo-Instruct model \citep{openai2023} to generate concise document summaries. These summaries, approximately 20-40 words in length, are designed to capture the core points and main themes of each document. This conciseness helps ensure that the summaries are broad enough to represent a wider range of documents within each topic, facilitating the creation of more inclusive and representative topic names. We then input these summaries into the BERTopic model \citep{grootendorst2022}, which generates the top keywords representing each topic. Both the document summaries and the generated keywords are saved and utilized in the subsequent topic labeling approaches.

\subsection{Topic Labeling}

We explore four distinct topic labeling approaches, each utilizing keywords and/or document summaries in different ways. Once the relevant information is collected for each topic, it is processed by the GPT-3.5-Turbo-Instruct model \citep{openai2023}. Using a persona-based prompt, the model generates a concise topic name, ranging from two to five words, that best represents the provided information. An overview of these approaches is also illustrated in Figure 1.

\subsubsection{Approach 1}

This approach ranks the top ten documents (summaries) for each topic based on their overlap with the topic's top ten keywords. By evaluating the direct overlap between the words in the documents and the key terms defining the topic, this method selects documents that most closely align with the core topic. The goal is to identify documents that best encapsulate the topic's essential elements, as defined by the keywords.

\subsubsection{Approach 2}

For each topic, this approach constructs a matrix of dimensions \textit{N×N} (where \textit{N} is the number of documents in the topic) to store pairwise cosine similarities between all documents. These similarities are computed using TF-IDF representations. After calculating the cosine similarities, a total similarity score for each document is obtained by summing the values in its corresponding row of the matrix. The top ten documents are then selected based on these scores. This approach aims to identify documents that are most closely related to other documents within the topic, capturing a more holistic view of the document content.

\subsubsection{Approach 3}

In this approach, topics are further divided into subtopics using the BERTopic model \citep{grootendorst2022}. From the largest subtopic within each main topic, ten documents are randomly sampled. This method provides a finer distinction within the overarching topic and focuses on the most prominent subtopic to represent the main theme. Although this approach might not fully capture the entire topic, it highlights the dominant theme within the topic.

\subsubsection{Approach 4}

Similar to Approach 3, this method involves dividing topics into subtopics using BERTopic. However, instead of focusing on the largest subtopic, this approach aims to achieve greater diversity by sampling one document from each subtopic. If a topic contains more than ten subtopics, ten documents are randomly sampled from the aggregated documents of all subtopics to ensure both a representative sample and consistency with the output of the other approaches. This approach seeks to incorporate a broader range of perspectives within the topic, balancing diversity and representation.

\section{Experimental Setup}

\subsection{Datasets}
We use two datasets to validate the performance of our methods: BBC News and 20 Newsgroups. 

The BBC News dataset, accessed from Kaggle, contains 2,225 documents across five main areas from the BBC News website, spanning from 2004 to 2005 \citep{greene2006}. The 20 Newsgroups dataset, accessed from the Scikit-Learn library, contains around 18,000 documents spanning 20 categories \citep{lang1995} These datasets were chosen for their diversity and variation in terms of document length and content.

\subsection{Evaluation}

A novel evaluation metric is proposed to assess the effectiveness of the topic labeling approaches by measuring how semantically representative a label is for the set of documents it describes. First, embeddings for both the topic labels and documents are generated using the Sentence-BERT model \citep{reimers2019}. The pairwise semantic coherence, measured through cosine similarity between the label and each document, is then calculated and averaged at the topic level. Finally, a weighted average of the topic-level scores is computed, with weights based on the number of documents per topic, ensuring that larger or broader topics do not disproportionately influence the overall metric. This metric is relative, with higher values indicating better label quality.

\subsection{Experimental Settings}

We run the BERTopic model \citep{grootendorst2022} once for each dataset. We access and save the top 10 keywords as well as the documents in each topic, which are then used in the four approaches to ultimately input them in the LLM.

\begin{table*}
    \centering
    \small
    \scalebox{1.29}{
        \begin{tabular}{|C{1in}|C{1in}|C{1in}|C{1in}|} \hline   
             \textbf{Approach 1} &  \textbf{Approach 2} &  \textbf{Approach 3} & \textbf{Approach 4} \\\hline
             \multicolumn{4}{|c|}{BBC} \\ \hline
             {Energy Partnership with Russia} &  {Yukos' Legal Battle for Survival} &  {Tax Troubles in Gazprom's Nuclear Division} & {Tax Dispute Over Nuclear Exports}\\\hline
             {Rising from Disaster: Economic Impact of Natural Disasters} &  {Global Manufacturing Cost Reduction} &  {Legal Professionals Against Indefinite Detention} & {UK Economic Slowdown Forecasted}\\\hline
            {Clean Competition Initiative} &  {Sprinters' Drug Test Dilemma} &  {Doping Bans in Professional Tennis} & {Controversial Drug Test Results: Koubek's Appeal}\\\hline
        \multicolumn{4}{|c|}{20 Newsgroups} \\ \hline
        {Controversies Surrounding Encryption Technology} &  {Encryption and Government Control} &  {Crypto Chip Controversy Coalition} & {Global Privacy and Government Surveillance}\\\hline
             {Questioning Waco: Inconsistencies and Government Actions} &  {Waco Siege: Questioning Government Actions} &  {The Waco Siege: FBI and BATF Controversy} & {Government Press Targeting and Safety Concerns}\\\hline
            {Cancer Treatment and Side Effects} &  {Debating Health and Medical Understanding} &  {Health and Wellness: Treatment and Prevention} & {Effective Treatment: Medical Information and Education}\\\hline
        \end{tabular}
    }
    \caption{Topic Labeling Outputs for Various Approaches on the BBC and 20 Newsgroups Datasets}
    \label{tab:my_label}
\end{table*}

\section{Results and Discussion}
Upon generating the topic labels, we evaluated each of the four strategies using the metric proposed above. The results for both datasets are presented in Table 1. We also display the outputs generated from the four approaches for some of the topics for both datasets in Table 2.

As shown, Approach 3, which selects the document summaries from the largest subtopic, yields the highest performance for the BBC dataset, achieving a score of 0.1200—significantly higher than the other three approaches. This can be attributed to the well-defined categories in the BBC dataset, where subtopics will align closely with the main topics. The largest subtopic is likely to capture the core thematic essence of the topic, whereas other approaches, which focus more on diversity or finer details, may introduce noise and weaken overall topic representation. In contrast, Approach 2 yields the best performance for the 20 Newsgroup dataset, followed by Approach 3. The high similarity between categories in this dataset, with overlapping words and concepts, makes it challenging to define topics based solely on dominant keywords or individual documents. By leveraging TF-IDF to identify the most representative document summaries, Approach 2 effectively captures the underlying structure of each topic, resulting in more informative and coherent labels.

The BBC dataset exhibits greater sensitivity to the choice of approach, as reflected in the relatively larger differences in performance across the four strategies. This suggests that certain selection methods are more effective due to the dataset’s well-defined categories, where subtopics align closely with the main topics. In contrast, the 20 Newsgroup dataset, with its overlapping and interrelated categories, shows less fluctuation across approaches. The lower topic separability likely contributes to the strong performance of Approach 2, which prioritizes semantic coherence. 

These findings highlight the importance of selecting topic-labeling strategies based on dataset characteristics and emphasize the need for experimentation to determine the most suitable approach. However, Approach 3 has demonstrated strong performance across both datasets, suggesting it may serve as a reliable starting point for new datasets. In contrast, Approach 4 has consistently underperformed, indicating that prioritizing diversity in this manner may not be the most effective strategy for generating coherent topic labels.

\section{Limitations}

There are several limitations to our approach. First, the effectiveness of our method heavily relies on the quality of the summaries generated by the LLM in use. In this study, we employed GPT-3.5-Turbo-Instruct \citep{openai2023}, which is a closed-source model, limiting our control and insight into its inner workings. Additionally, the model's context window size poses a limitation, as it may lead to potential information loss. Although this was not a significant issue in this study, it could become problematic for datasets with larger documents. Our approach is also not suitable for datasets with very short documents, such as tweets, where summarization is neither feasible nor effective. Attempting to summarize such brief texts may lead to worse results.

\section{Conclusion and Future Work}

We propose a method to enhance topic labeling by leveraging large language models (LLMs) to generate concise and meaningful labels for topic clusters. Our study proposes four different document selection strategies, each emphasizing different aspects, such as dominant themes or diversity, to optimize the input for the LLM. Observing the lack of a quantitative metric for evaluating topic labeling approaches, we also introduce a metric that measures how semantically representative a label is of its corresponding topic documents. Our experimental results provide insights into the effectiveness of each approach, revealing that Approach 3 generally yields the best results. Based on our findings, we recommend starting with Approach 3 and adapting other strategies based on dataset characteristics. 

Future research should explore additional ways in which LLMs can refine topic labels. Further work is also needed to develop and validate new quantitative metrics for assessing topic labeling performance, considering aspects such as representativeness and diversity. Additionally, the effectiveness of the proposed metric should be verified through human evaluations to ensure alignment with qualitative assessments.

To assess the broader applicability of our approach, it should be tested on a more diverse set of datasets, particularly those beyond news documents. Since the effectiveness of our method depends on the quality of the LLM used, advancements in LLM technology are expected to enhance topic labeling outcomes. 

\bibliographystyle{acl_natbib}
\bibliography{custom}

\begin{thebibliography}{13}
\providecommand{\natexlab}[1]{#1}

\bibitem[{Aletras and Stevenson(2014)}]{aletras2014}
Nikolaos Aletras and Mark Stevenson. 2014.
\newblock Labelling topics using unsupervised graph-based methods.
\newblock In \emph{Proceedings of the 52nd Annual Meeting of the Association for Computational Linguistics (Volume 2: Short Papers)}, pages 631--636. ACM Press.

\bibitem[{Alokaili et~al.(2020)Alokaili, Aletras, and Stevenson}]{alokaili2020}
Areej Alokaili, Nikolaos Aletras, and Mark Stevenson. 2020.
\newblock Automatic generation of topic labels.
\newblock \emph{arXiv preprint arXiv:2006.00127}.

\bibitem[{Greene and Cunningham(2006)}]{greene2006}
Derek Greene and Pádraig Cunningham. 2006.
\newblock Practical solutions to the problem of diagonal dominance in kernel document clustering.
\newblock In \emph{Proceedings of the 23rd International Conference on Machine Learning (ICML'06)}, pages 377--384. ACM Press.

\bibitem[{Grootendorst(2022)}]{grootendorst2022}
Maarten Grootendorst. 2022.
\newblock Bertopic: Neural topic modeling with a class-based tf-idf procedure.
\newblock \emph{arXiv preprint arXiv:2203.05794}.

\bibitem[{Kozlowski et~al.(2024)Kozlowski, Pradier, and Benz}]{kozlowski2024}
Diego Kozlowski, Carolina Pradier, and Pierre Benz. 2024.
\newblock Generative ai for automatic topic labelling.
\newblock \emph{arXiv preprint arXiv:2408.07003}.

\bibitem[{Lang(1995)}]{lang1995}
Ken Lang. 1995.
\newblock Newsweeder: Learning to filter netnews.
\newblock In \emph{Machine Learning Proceedings 1995}, pages 331--339. Elsevier.

\bibitem[{Lau et~al.(2011)Lau, Grieser, Newman, and Baldwinn}]{lau2011}
Jey~Han Lau, Karl Grieser, David Newman, and Timothy Baldwinn. 2011.
\newblock Automatic labelling of topic models.
\newblock In \emph{Proceedings of the 49th Annual Meeting of the Association for Computational Linguistics: Human Language Technologies}, page 1536–1545. ACM Press.

\bibitem[{Mei et~al.(2007)Mei, Shen, and Zhai}]{mei2007}
Qiaozhu Mei, Xuehua Shen, and ChengXiang Zhai. 2007.
\newblock Automatic labeling of multinomial topic models.
\newblock In \emph{Proceedings of the 13th ACM SIGKDD International Conference on Knowledge Discovery and Data Mining}, pages 490--499. ACM Press.

\bibitem[{OpenAI(2023)}]{openai2023}
OpenAI. 2023.
\newblock {GPT}-3.5 turbo.
\newblock Retrieved from \url{https://platform.openai.com/docs/models/gpt-3-5}.

\bibitem[{Reimers and Gurevych(2019)}]{reimers2019}
Nils Reimers and Iryna Gurevych. 2019.
\newblock Sentencebert: Sentence embeddings using siamese bert-networks.
\newblock In \emph{Proceedings of the 2019 Conference on Empirical Methods in Natural Language Processing}. Association for Computational Linguistics.

\bibitem[{Rijcken et~al.(2023)Rijcken, Scheepers, Kalliopi~Zervanou, Mosteiro, and Kaymak}]{rijcken2023}
Emil Rijcken, Floortje Scheepers, Marco~Spruit Kalliopi~Zervanou, Pablo Mosteiro, and Uzay Kaymak. 2023.
\newblock Towards interpreting topic models with chatgpt.
\newblock In \emph{The 20th World Congress of the International Fuzzy Systems Association}.

\bibitem[{Wan and Wang(2016)}]{wan2016}
Xiaojun Wan and Tianming Wang. 2016.
\newblock Automatic labeling of topic models using text summaries.
\newblock In \emph{Proceedings of the 54th Annual Meeting of the Association for Computational Linguistics (Volume 1: Long Papers)}, page 2297–2305. ACM Press.

\bibitem[{Zhang et~al.(2020)Zhang, Kishore, Wu, Weinberger, and Artzi}]{zhang2020}
Tianyi Zhang, Varsha Kishore, Felix Wu, Kilian~Q. Weinberger, and Yoav Artzi. 2020.
\newblock Bertscore: Evaluating text generation with bert.
\newblock \emph{arXiv preprint arXiv:1904.09675}.

\end{thebibliography}

\end{document}